\documentclass[%
reprint,
 amsmath,amssymb,
 aps,
]{revtex4-1}

\usepackage{graphicx}% Include figure files
\usepackage{dcolumn}% Align table columns on decimal point
\usepackage{bm}% bold math
\usepackage[export]{adjustbox}
\usepackage{subcaption}
\usepackage{siunitx}
\usepackage{hyperref}
\usepackage{natbib}

\begin{document}

\title{Frequency-Stabilized Deep-UV Laser at 243.1 nm with 1.4 W output power}

\author{Z. Burkley}%
\email{zakary.burkley@colostate.edu}
\author{A.D. Brandt}%
\author{C. Rasor}%
\author{S.F. Cooper}%
\author{D.C. Yost}%

\affiliation{%
Department of Physics, Colorado State University, Fort Collins, CO 80523\\
}%

%% To be edited by editor
% \dates{Compiled \today}

%\ociscodes{(140.3615) Lasers, ytterbium; (060.2320) Fiber optics amplifiers and oscillators; (140.3515) Lasers, frequency doubled; (140.3610) Lasers, ultraviolet; (300.6540) Spectroscopy, ultraviolet.}

%% To be edited by editor
% \doi{\url{http://dx.doi.org/10.1364/XX.XX.XXXXXX}}
\date{\today}

\begin{abstract}
We demonstrate a 1.4 W continuous wavelength (CW) laser at 243.1 nm. The radiation is generated through frequency quadrupling the output of a ytterbium-doped fiber amplifier system which produces $>$ 10 W of CW power at 972.5 nm. We demonstrate absolute frequency control by locking the laser to an optical frequency comb and exciting the 1\textit{S}--2\textit{S} transition in atomic hydrogen. This frequency-stabilized, high-power deep-UV laser should be of significant interest for precision spectroscopy of simple and exotic atoms, two-photon laser cooling of hydrogen, and Raman spectroscopy. 
\end{abstract}

\pacs{Valid PACS appear here}% PACS, the Physics and Astronomy
\maketitle

\section{Introduction}

\noindent
High-power fiber lasers from 1-1.2 $\mu$m have flourished as a result of advances in ytterbium(Yb)-doped fiber amplifiers \cite{richardson2010}. In addition, Yb-fiber lasers between 970--980 (97x) nm have also been developed --- a notable accomplishment due to gain competition in the $>$ 1 $\mu$m spectral region \cite{boullet2008, roeser2008}. These systems initially lacked sufficiently narrow spectral bandwidth for efficient harmonic generation, motivating further development since there is significant interest in frequency doubling and quadrupling these sources to produce coherent blue radiation and deep-UV radiation.  For example, spectroscopy on positronium and muonium would benefit from high-power 486 nm and 244 nm radiation, respectively. While these exotic atoms are ideal for testing bound state QED \cite{karshenboim2005}, the measurements face low detection rates \cite{cooke2015, Crivelli2018} and would benefit from increased laser power.  Additionally, a high powered 243 nm source allows for an increased interaction volume for 1\textit{S}--2\textit{S} spectroscopy on hydrogen and anti-hydrogen, which reduces uncertainities caused by transit-time broadening and statistics  \cite{parthey2011, ahmadi2017}. Furthermore, enhancement of high-power 243 nm radiation within a cavity could enable two-photon laser cooling of hydrogen \cite{zehnle2001, kielpinski2006, cooper2018}.\\

\indent Coherent, high-power deep-UV lasers at 243--244 nm also have utility outside of the metrology of simple atomic systems. In Raman spectroscopy, the signal intensity follows a $\lambda^{-4}$ dependence, motivating the use of UV lasers. Due to the relatively low cross sections of Raman processes, high-power UV sources could help in detection of trace gases, which could assist in identification of explosive materials \cite{nagli2008}. While high-power 266 nm lasers are available \cite{sakuma2004high}, a 243--244 nm laser can be preferable in some cases as the flourescence background in Raman spectroscopy diminishes rapidly below 250 nm \cite{asher1993}. The 243--244 nm wavelength range also matches the semiconductor industry's standard KrF photoresists. Therefore, this laser system offers an alternative to the argon-ion lasers often used in laser interference lithography \cite{seo2014}.\\

\indent Motivated by improving spectroscopy of simple and exotic atoms, laser cooling hydrogen, and the general utility of a high-power deep-UV laser, we previously developed a Yb-fiber amplifier tunable from 972--976 nm \cite{burkley2017, cooper2018}. This system was able to generate $\approx$ 6 W of 972--976 nm radiation limited by the available pump power (50 W).  The linewidth of this source was narrow enough such that $\approx$ 0.5 W of CW, deep-UV radiation from 243--244 nm could be generated through intracavity harmonic generation. While our previous work suggested that this Yb-fiber laser could produce additional power if more pump power was available, it was unclear if the laser amplifier and frequency doubling stages were truly power scalable. Furthermore, the previous system did not use a frequency-stabilized source, and no in-depth studies of the linewidth were made --- both critical requirements for a spectroscopy laser.  Recently, a Yb-fiber laser system similar to ours was investigated that generated 10 W of narrow linewidth ($<$ 100 kHz), linear polarized CW power at 976 nm \cite{wu2018}. While harmonic generation was not implemented in this work, design of this laser system was strongly motivated at the prospect of developing Watt-level, coherent deep-UV sources. \\

\indent In this letter, we investigate both the power scalability and frequency control of our deep-UV laser system. We demonstrate that with additional pump power (120 W), our Yb-fiber amplifier produces 10.8 W of frequency-stabilized 972.5 nm radiation. Through two successive frequency doubling stages the system is capable of generating 4.2 W at 486.3 nm and 1.4 W at 243.1 nm. To the best of our knowledge, this is the highest power deep-UV, CW laser below 266 nm, and builds onto significant advancements in high-power frequency-quadrupled systems in recent years \cite{kaneda2016, zhao2017}. Finally, we estimate the linewidth of the radiation at 243.1 nm is $<$ 10 kHz and demonstrate the frequency control of this deep-UV laser through excitation of the 1\textit{S}--2\textit{S} transition in atomic hydrogen.

\section{Laser Design and Performance}

\begin{figure} [h]
\centering\includegraphics[width = \linewidth]{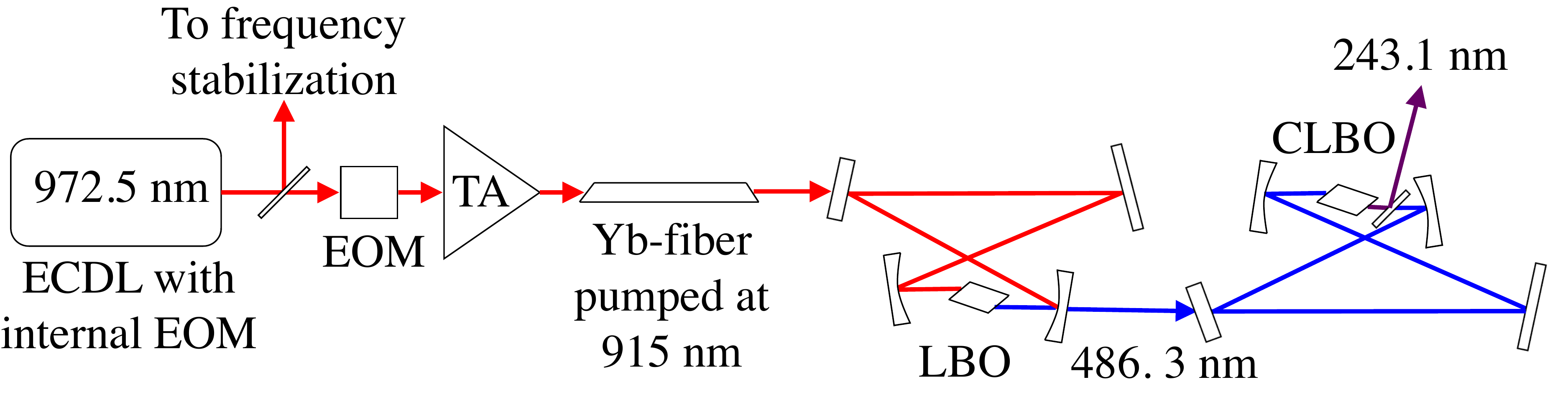}
\caption{Simplified design of the Yb-fiber amplifier and resonant doubling stages. The master oscillator at 972.5 nm is an extended cavity diode laser (ECDL). A small portion of the 972.5 light is used to actively stabilize the frequency of this oscillator. This 972.5 nm light is then amplified through a tapered amplifier (TA) and Yb-doped amplifier, followed by frequency quadrupling through successive resonant doubling stages. The EOM directly before the TA provides the necessary modulation for locking these stages on resonance using the Pound-Drever Hall method. A more detailed discussion of this system is given in \cite{burkley2017}.}
  \label{fig:design}
\end{figure}

Our laser system, shown in Fig. \ref{fig:design} and described in more detail in \cite{burkley2017}, starts with a frequency-stabilized extended cavity diode laser (ECDL), which operates at 972.5 nm. The output of the ECDL is then amplified in two stages: first with a commercial tapered amplifier, and then with a double-clad Yb-doped fiber amplifier. This amplified light is then frequency quadrupled to generate the desired 243.1 nm light.

Amplification of the ECDL output at 972.5 nm with the tapered amplifier generates 3 W of power, with $\approx$ 2.4 W in a TEM00 Gaussian mode. This light is then coupled to the Yb-doped fiber amplifier. While Yb-fiber is generally used for generation and amplification of 1030 nm light, it is also possible to achieve gain at 97x nm given sufficient population inversion. At 976 nm, which corresponds to the peak emission cross section of ytterbium, generating gain requires 50\% population inversion. Below 976 nm, the emission cross section drops, requiring even larger population inversions to generate gain at these wavelengths. This is problematic as the gain at the more favorable spectral regions at 1030 nm and 976 nm will reduce the population inversion and degrade the Yb-fiber's amplification at 972.5 nm. We combat parasitic gain at these more favorable lasing spectral regions in several ways. First, we use a fiber with a large core/cladding ratio \cite{boullet2008, nilsson1998, roeser2008}. Second, the fiber amplifier is kept relatively short ($\approx$ 10 cm). Third, by angle polishing the ends of the fiber, we reduce the back-reflections at the air/fiber interface. Finally, we heavily seed the fiber amplifier with 972.5 nm light as described above and aggressively filter unwanted wavelengths.

\begin{figure} [h]
\centering\includegraphics[width = 0.85\linewidth]{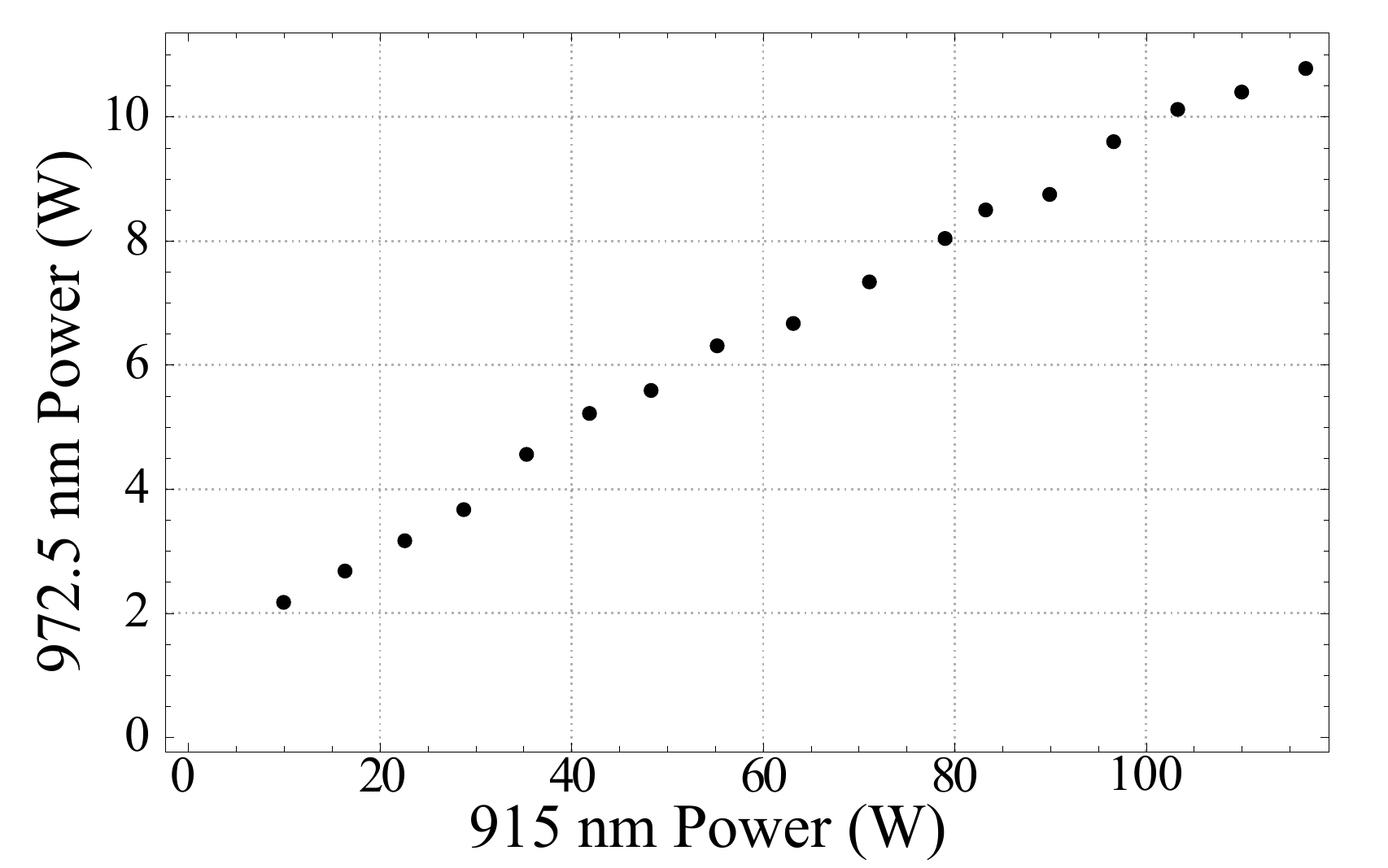}
\caption{Measured 972.5 nm Yb-fiber output power vs. launched 915 nm pump power. With zero pump power, the 972.5 nm Yb-fiber output power does not pass through zero as the seed power at 972.5 nm is large enough to invert the medium to a point of semi-transparency.}
  \label{fig:IR}
\end{figure}

\begin{figure} [h!]
\centering\includegraphics[width = 0.85\linewidth]{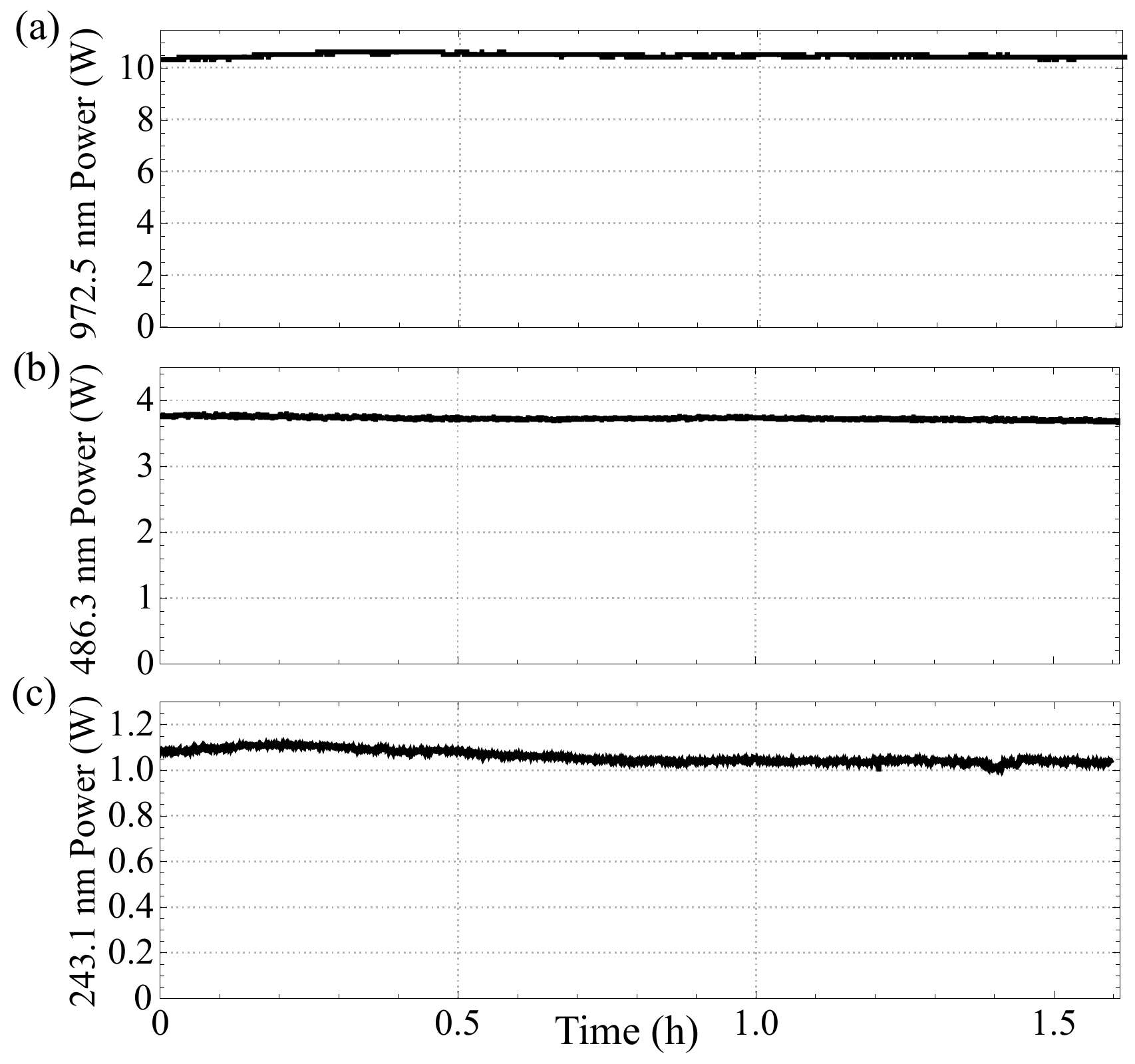}
\caption{The power stability of the amplified 972.5 nm radiation and successive harmonics at 486.3 nm and 243.1 nm. The data for these three curves was acquired at different times. To improve thermal stability, the powers shown here are derated from the maximum powers the laser system can generate at these three wavelengths.}
  \label{fig:stability}
\end{figure}

A plot of the Yb-fiber amplified 972.5 nm power versus launched 915 nm pump power is shown in Fig. \ref{fig:IR}. We are able to generate up to 10.8 W of 972.5 nm radiation with 120 W of launched pump power. There is some concern that high inversions can lead to photodarkening, especially as the system is power scaled. However, our commercially available fiber is co-doped with phosphorous, which mitigates this effect \cite{engholm2008}. As seen in Fig. \ref{fig:stability}(a), stable IR power can be maintained at $>$ 10 W over time periods greater than 1 hour. We have run the system for a combined duration greater than 20 hours at these high powers, and see no evidence of degradation from photodarkening. Overall, we believe this system is still power scalable with more pump power. However, the free-space coupling of both the seed and pump light \cite{burkley2017} requires significant thermal management of the fiber mount and coupling lens mounts as Fresnel reflections off the fiber tips and scatter off the optics heats these components.  We currently address these issues using radiation shields and water cooling.

\begin{figure} [h!]
\centering\includegraphics[width = 0.85\linewidth]{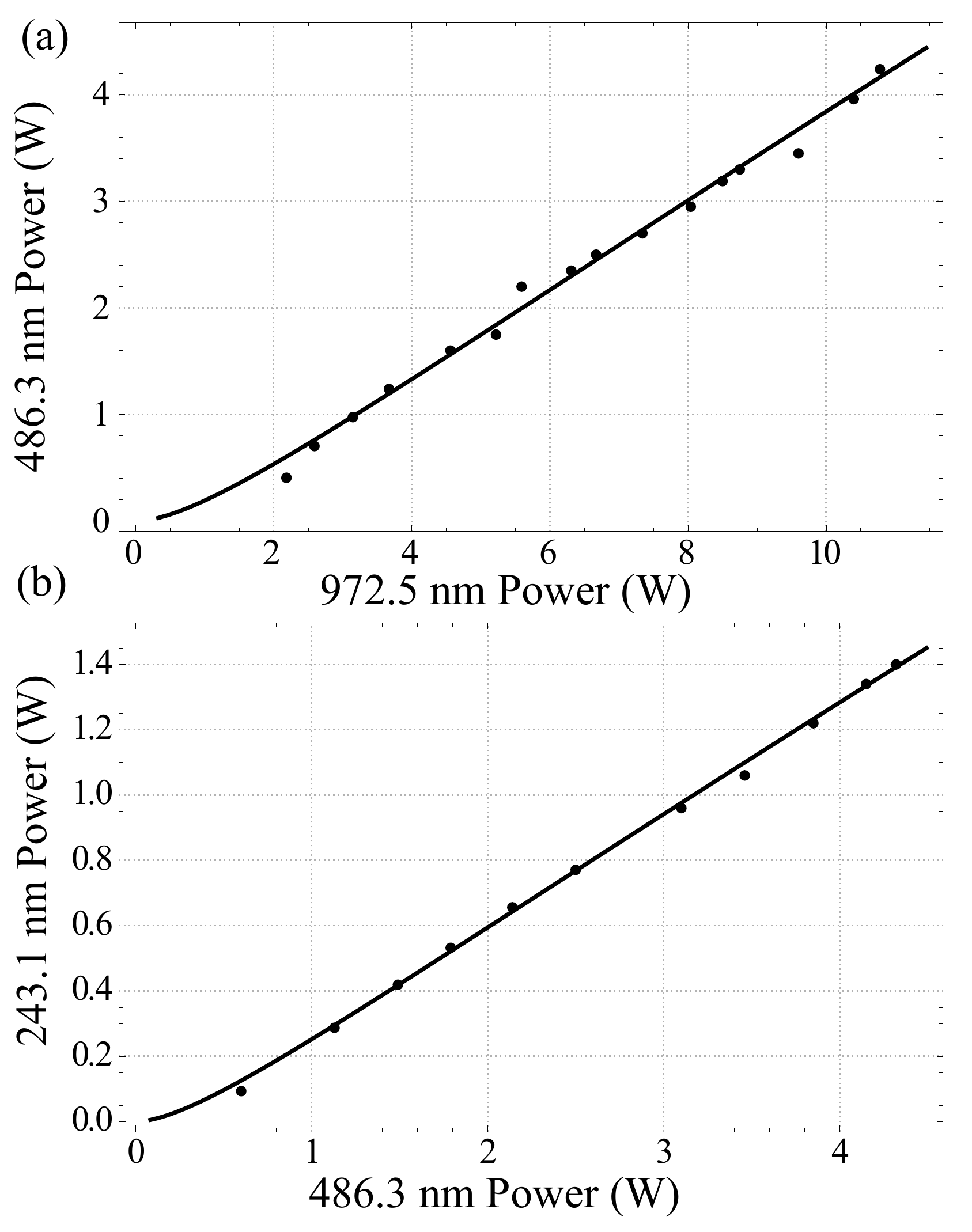}
\caption{Frequency doubling results. a) Measured 486.3 nm power vs. input 972.5 nm power using LBO as the nonlinear medium b) Measured 243.1 nm power vs. input 486.3 nm power using CLBO as the nonlinear medium. The theoretical fits for harmonic conversion in a) and b) follow the methods presented in \cite{polzik1991}.}
  \label{fig:doub}
\end{figure}

The amplified radiation is then sent through successive resonant doubling stages, which are described in more detail in \cite{burkley2017}, generating 486.3 nm and 243.1 nm, respectively. The first resonant doubling stage uses a 25 mm long LBO crystal, a 3\% input coupler, and 150 mm radius of curvature mirrors that generate a beam waist of $\approx$ 50 $\mu$m within the crystal. This relatively large beam waist was chosen to make the system more robust when operating at high output power. A plot of 486.3 nm cavity output power versus input 972.5 nm power from this first doubling stage is shown in Fig. \ref{fig:doub}(a). As can be seen, we obtain 4.2 W of 486.3 nm with 10.8 W of 972.5 nm input power. 

For the second resonant doubling stage, we use a 10 mm long CLBO crystal, 2.5\% input coupler, and 200 mm radius of curvature mirrors that generate a 44 $\mu$m beam waist within the crystal. Since CLBO is hygroscopic, and absorbed moisture negatively impacts its performance \cite{kawamura2009}, we keep the crystal at an elevated temperature of 150 $^{\circ}$C with a continual dry nitrogen purge. A plot of outcoupled 243.1 nm power versus input 486.3 power is shown in Fig. \ref{fig:doub}(b). With 4.2 W of 486.3 nm power, we are able to generate 1.4 W of 243.1 nm power.

Due to the high UV powers generated, UV-induced degradation of the CLBO crystal is a concern. While 5 W of 266 nm, 120 mW of 244 nm, and 140 mW of 198.5 nm CW radiation have been maintained for several hours using CLBO without evidence of degradation \cite{sakuma2004high, kaneda2008, 2004sakuma}, studies observing degradation in CLBO on long time scales are limited and often restricted to pulsed laser systems. Using a 266 nm pulsed laser, it was shown that a UV-induced refractive index change can cause degradation in CLBO at peak power densities orders of magnitude lower than the bulk-induced damage threshold \cite{takachiho2014}. This degradation is evidenced by a decrease in UV transmittance of the crystal over time, with a rate that increases with UV peak power density. Therefore, we use a beam waist $\approx$ 2 times larger than the optimal Boyd-Kleinmann parameter in our CLBO crystal. With this, we have generated greater than 1 W of 243.1 nm radiation for a combined duration greater than 20 hours on the same spot of the crystal without signs of UV-induced refractive index change degradation. Although our peak power density is significantly less than the pulsed system used in \cite{takachiho2014}, our CW intensity is comparable to the average intensities for which they saw degradation on a sub-hour timescale.

While we did not observe degradation of the CLBO crystal, we have seen evidence of non-linear absorption in the crystal resulting in self-heating and harmonic phase mismatch. This was observed in the output 243.1 nm beam which would oscillate around the optimal phase-matching condition over a few second period. We were able to mitigate this effect by adjusting the phase-matching temperature slightly. While the system was capable of producing 1.4 W of output power at 243.1 nm stably over tens of minutes, the thermal stability improved if the output was marginally derated.  As shown in Fig. \ref{fig:stability}(c), we observed that we could maintain more than 1 W of 243.1 nm for over 1 hour without observing this instability due to self-heating.  

\begin{figure} [h!]
\centering\includegraphics[width = 0.8\linewidth]{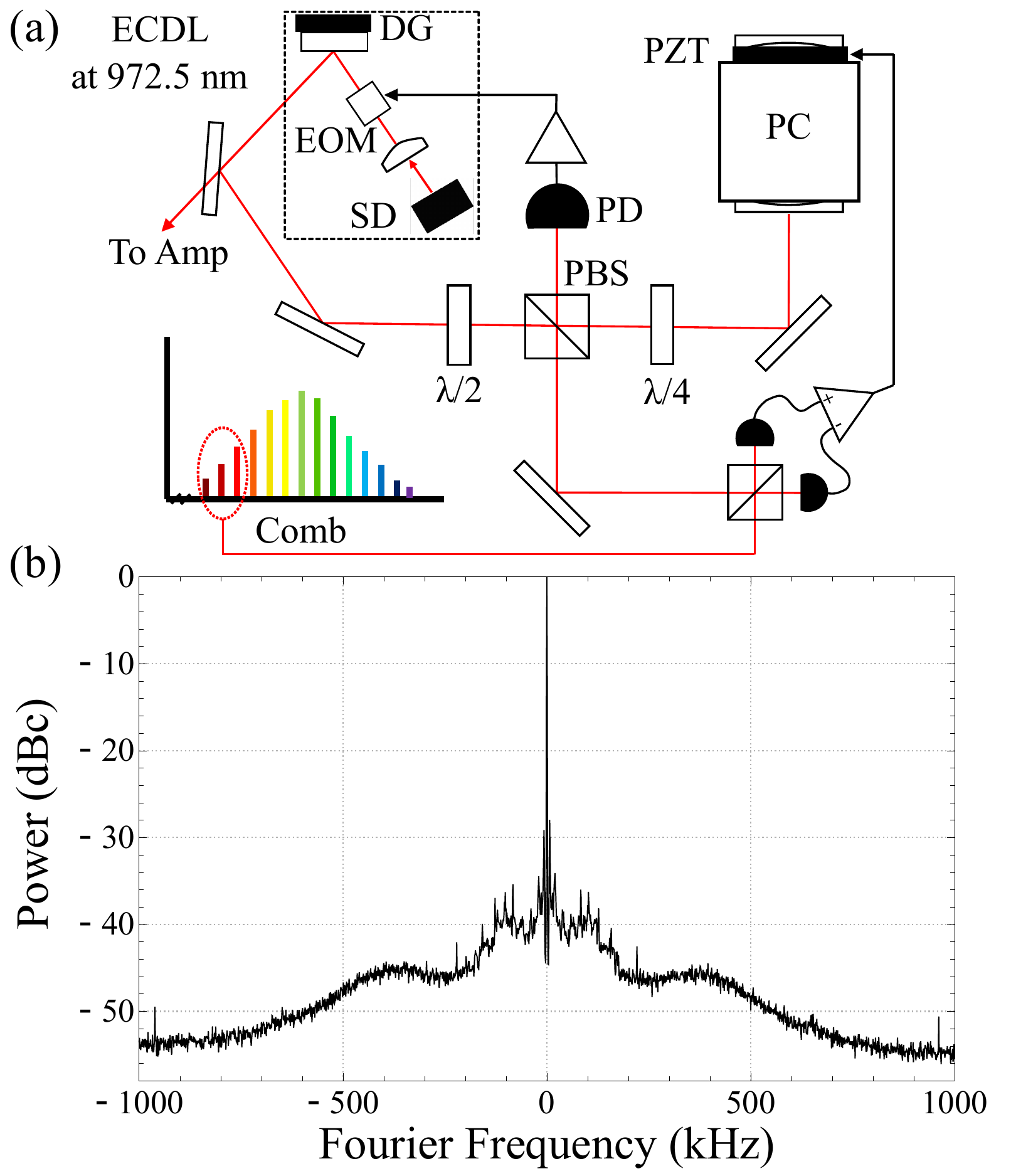}
  \caption{(a) Schematic of the 972.5 nm locking system. SD: seed diode, EOM: electro-optic modulator, DG: diffraction grating, ECDL: extended cavity diode laser, $\lambda/2$: half-wave plate, $\lambda /4$: quarter-wave plate, PBS: polarization beam-splitter, PD: photodiode, PC: pre-stabilization cavity, PZT: piezo-electric transducer. Frequency stabilization of the master oscillator at 972.5 nm involves simultaneous locking to a pre-stabilization cavity and an optical frequency comb. Feedback from the PC to the master oscillator is done with an EOM internal to the ECDL, whereas a PZT on the PC is used to lock the frequency of the 972.5 nm radiation to the frequency comb. (b) RF spectrum of the coherent beat note between the master oscillator at 972.5 nm and frequency comb taken at a resolution bandwidth of 1 kHz.}
  \label{fig:beatnote}
\end{figure}

\section{Frequency Control and 1S-2S Excitation}

To perform precision spectroscopy on simple and exotic atomic species, it is imperative that the laser frequency is well-controlled and determined. Additionally, control over the absolute frequency of the 243.1 nm radiation is critical to perform two-photon Doppler-cooling on atomic hydrogen beams. As shown in Fig. \ref{fig:beatnote}(a), we stabilize the 972.5 nm oscillator by first locking to a pre-stablization reference cavity, and then by adjusting the reference cavity's length to lock the oscillator to a coherent, GPS referenced frequency comb \cite{brandt2017}. The beat note between the oscillator and the frequency comb is shown in Fig. \ref{fig:beatnote}(b). Based on this coherent beat note and the in-loop error signal with the pre-stablilization cavity, we estimate that $\approx$ 95\% of our optical power at 972.5 nm is in the coherent carrier. While the percentage of power in the carrier is degraded through frequency quadrupling \cite{alnis2008}, we estimate the linewidth at 243.1 nm is $<$ 10 kHz. This is sufficient for spectroscopy on muonium and positronium due to the short lifetime of the species; as well as two-photon laser cooling of hydrogen since the transition would be broadened substantially by coupling the 2S state to the 2P state for quenching \cite{zehnle2001, kielpinski2006}. However, for state-of-the-art spectroscopy on the 1\textit{S}--2\textit{S} transition \cite{alnis2008, parthey2011}, the linewidth of our source would need to be reduced further. 

\begin{figure} [h!]
\centering\includegraphics[width=0.8\linewidth]{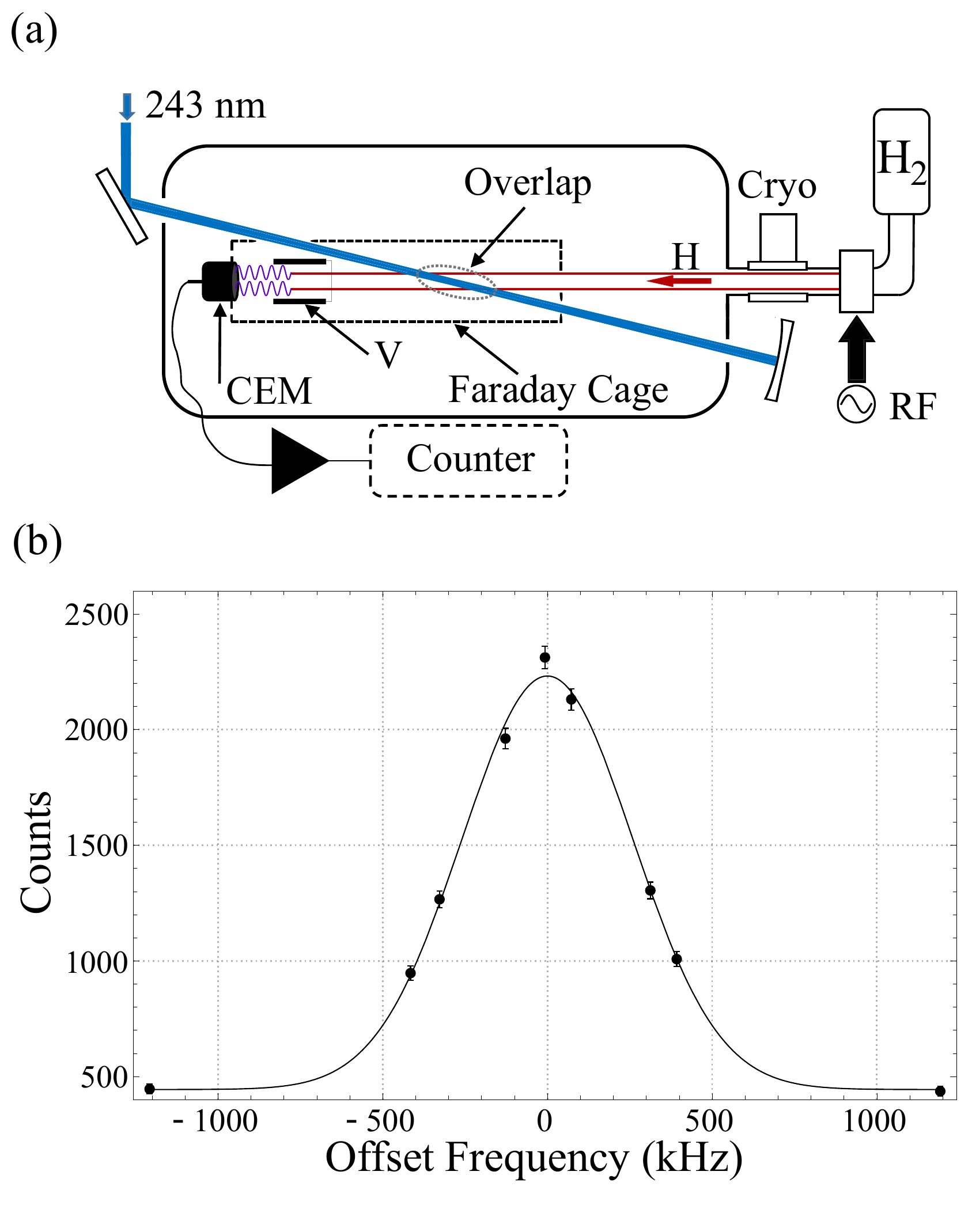}
  \caption{(a) Schematic of the hydrogen 1\textit{S}--2\textit{S} excitation. CEM: channel electron multiplier, V: quench electrodes, RF: microwave discharge, H$_{2}$: molecular  hydrogen, Cryo: cryostat, Counter: frequency counter. The UV and atomic beam overlap in a Faraday cage to prevent quenching from stray fields. (b) Measured lineshape of the 1\textit{S}--2\textit{S} transition.}
  \label{fig:lineshape}
\end{figure}

As a demonstration of the absolute frequency control and stability of the UV light, we excited the 1\textit{S}--2\textit{S} transition in hydrogen. We used only a portion of the available power ($\approx$ 200 mW) and a beam waist of 80 $\mu$m. As shown in Fig. \ref{fig:lineshape}, this radiation was overlapped with a 50 K atomic hydrogen beam at an angle of 6$^{\circ}$, exciting some population into the metastable state. These metastable atoms were then quenched by an electric field, and the emitted Lyman-$\alpha$ photons were counted by a channel electron multiplier. The recovered 1\textit{S}--2\textit{S} lineshape is Gaussian, with a FWHM of 850 kHz, which is in excellent agreement with our estimate for transit-time broadening. 

\section{Conclusion}

To conclude, we have demonstrated a Yb-fiber amplifier generating up to 10.8 W of 972.5 nm light. This increase in IR power directly led to larger output powers of the frequency doubled and quadrupled light, with the 243.1 nm light reaching powers of 1.4 W. Operating the system at high powers for $>$ 20 hours has not yet shown degradation of the Yb-fiber amplifier, LBO or CLBO crystals, and we have demonstrated robust continuous high-power operation over 1 hour timescales.

In addition, we have also investigated the frequency stability and linewidth of the laser source, demonstrating excitation of the 1\textit{S}--2\textit{S} transition in hydrogen. We currently estimate that the linewidth at 243.1 nm is $<$ 10 kHz. This frequency-stabilized deep UV laser with high output power represents a powerful tool for precision spectroscopic measurements in hydrogen, and spectroscopy on exotic species, such as muonium, positronium, and antihydrogen. Beyond a valuable tool for the metrology on these simple and exotic atoms, we hope the technology used in this laser system can be applied to UV laser sources in general to improve their versatility, power scalability, and coherence properties.

\section{Acknowledgments}

We gratefully acknowledge useful discussions with Paolo Crivelli, Jacob Roberts, and Michael Morrison. Funding is provided from the NSF under Award \# 1654425. 

% Bibliography
\bibliography{references}

% Full bibliography added automatically for Optics Letters submissions; the following line will simply be ignored if submitting to other journals.
% Note that this extra page will not count against page length
%\bibliographyfullrefs{references}

% Please include bios and photos of all authors for aop articles

\end{document}